\documentstyle[aps,epsf,rotate,multicol]{revtex}

\begin{document}
\draft

\title{Statistical Properties of Share Volume Traded in Financial Markets}

\author{Parameswaran Gopikrishnan$^{1}$, Vasiliki Plerou$^{1,2}$,
Xavier Gabaix$^{3}$ and H. Eugene Stanley$^{1}$}

\address{$^{1}$ Center for Polymer Studies and Department of Physics,\\
Boston University Boston, Massachusetts 02215.\\
$^{2}$ Department of Physics, Boston College, Chestnut Hill, Massachusetts 
02164.\\
$^{3}$ Department of Economics, Massachusetts Institute of Technology, Cambridge, Massachusetts 02142.\\
}

\date{\today}

\maketitle

\begin{abstract} 

We quantitatively investigate the ideas behind the often-expressed
adage `it takes volume to move stock prices', and study the
statistical properties of the number of shares traded $Q_{\Delta t}$
for a given stock in a fixed time interval $\Delta t$. We analyze
transaction data for the largest 1000 stocks for the two-year period
1994-95, using a database that records every transaction for all
securities in three major US stock markets. We find that the
distribution $P(Q_{\Delta t})$ displays a power-law decay, and that
the time correlations in $Q_{\Delta t}$ display long-range
persistence. Further, we investigate the relation between $Q_{\Delta
t}$ and the number of transactions $N_{\Delta t}$ in a time interval
$\Delta t$, and find that the long-range correlations in $Q_{\Delta
t}$ are largely due to those of $N_{\Delta t}$. Our results are
consistent with the interpretation that the large equal-time
correlation previously found between $Q_{\Delta t}$ and the absolute
value of price change $\vert G_{\Delta t} \vert$ (related to
volatility) are largely due to $N_{\Delta t}$.
\end{abstract}
\begin{multicols}{2}

The distinctive statistical properties of financial time series are
increasingly attracting the interest of
physicists~\cite{Mantegna95}. In particular, several empirical studies
have determined the scale-invariant behavior of both the distribution
of price changes~\cite{Lux} and the long-range correlations in the
absolute values of price changes~\cite{Yanhui97}. It is a common
saying that `it takes volume to move stock prices'. This adage is
exemplified by the market crash of 19 October 1987, when the Dow Jones
Industrial Average dropped 22.6\% accompanied by an estimated $6\times
10^8$ shares that changed hands on the New York Stock Exchange
alone. Indeed, an important quantity that characterizes the dynamics
of price movements is the number of shares $Q_{\Delta t}$ traded
(share volume) in a time interval $\Delta t$. Accordingly, in this
paper we quantify the statistical properties of $Q_{\Delta t}$ and the
relation between $Q_{\Delta t}$ and the number of trades $N_{\Delta
t}$ in $\Delta t$. To this end, we select 1000 largest stocks from a
database~\cite{TAQ} recording all transactions for all US-stocks, and
analyze transaction data for each stock for the 2-year period
1994--95.

First, we consider the time series of $Q_{\Delta t}$ for one stock,
which shows large fluctuations that are strikingly non-Gaussian
[Fig.~1a]. Figure~1b shows, for each of four actively-traded stocks,
the probability distributions $P(Q_{\Delta t})$ which are consistent
with a power law decay,
\begin{equation}
P (Q_{\Delta t}) \sim {1 \over (Q_{\Delta t})^{1+\lambda}} \,.
\label{pdfQ}
\end{equation}
When we extend this analysis to the each of the 1000 stocks
[Fig.~1c,d], we obtain an average value for the exponent $\lambda =
1.7 \pm 0.1$, within the L\'evy stable domain $0 < \lambda < 2$.

We next analyze correlations in $Q_{\Delta t}$. We consider the family
of correlation functions $\langle [Q_{\Delta t}(t)]^a [Q_{\Delta
t}(t+\tau)]^a \rangle$, where the parameter $a$ ($< \lambda/2$) is
required to ensure that the correlation function is well
defined. Instead of analyzing the correlation function directly, we
apply detrended fluctuation analysis~\cite{Peng94}, which has been
successfully used to study long-range correlations in a wide range of
complex systems~\cite{Bunde}. We plot the detrended fluctuation
function $F(\tau)$ as a function of the time scale $\tau$. Absence of
long-range correlations would imply $F(\tau) \sim \tau^{0.5}$, whereas
$F(\tau) \sim \tau^{\delta}$ with $0.5 < \delta \leq 1$ implies
power-law decay of the correlation function,
\begin{equation}
\langle [Q_{\Delta t}(t)]^a [Q_{\Delta t}(t+\tau)]^a \rangle \sim 
\tau^{-\kappa}\,; \,\, [\kappa = 2 - 2\,\delta]\,.
\label{defQcorr}
\end{equation}
For the parameter $a=0.5$, we obtain the average value $\delta =
0.83\pm 0.02$ for the 1000 stocks [Fig.~2a,b]; so from
Eq.~(\ref{defQcorr}), $\kappa = 0.34 \pm 0.04$~\cite{diffa}.

To investigate the reasons for the observed power-law tails of
$P(Q_{\Delta t})$ and the long-range correlations in $Q_{\Delta t}$,
we first note that
\begin{equation}
Q_{\Delta t} \equiv \sum_{i=1}^{N_{\Delta t}} q_i \,,
\label{sumq}
\end{equation}
is the sum of the number of shares $q_i$ traded for all
$i=1,\dots,N_{\Delta t}$ transactions in $\Delta t$. Hence, we next
analyze the statistical properties of $q_i$. Figure~3a shows that the
distribution $P(q)$ for the same four stocks displays a power-law
decay $P(q) \sim 1/q^{1+\zeta}$.  When we extend this analysis to each
of the 1000 stocks, we obtain the average value $\zeta = 1.53 \pm
0.07$ [Fig.~3b].

Note that $\zeta$ is within the stable L\'evy domain $0~<~\zeta~<~2$,
suggesting that $P(q)$ is a positive (or one-sided) L\'evy stable
distribution~\cite{Levy,assymLevy}. Therefore, the reason why the
distribution $P(Q_{\Delta t})$ has similar asymptotic behavior to
$P(q)$, is that $P(q)$ is L\'evy stable, and $Q_{\Delta t}$ is related
to $q$ through Eq.~(\ref{sumq}). Indeed, our estimate of $\zeta$ is
comparable within error bounds to our estimate of $\lambda$. We also
investigate if the $q_i$ are correlated in ``transaction time'',
defined by $i$, and we find only ``weak'' correlations (the analog of
$\delta$ has a value $=0.57\pm 0.04$, close to $0.5$).

To confirm that $P(q)$ is L\'evy stable, we also examine the behavior
of $Q_n \equiv \sum_{i=1}^{n} q_i$.  We first analyze the asymptotic
behavior of $P(Q_n)$ for increasing $n$.  For a L\'evy stable
distribution, $n^{1/\zeta} \, P([Q_n - \langle Q_n \rangle] /
n^{1/\zeta})$ should have the same functional form as $P(q)$, where
$\langle Q_n \rangle=n\,\langle q \rangle$ and $\langle \dots \rangle$
denotes average values. Figure 4a shows that the distribution $P(Q_n)$
retains its asymptotic behavior for a range of $n$ --- consistent with
a L\'evy stable distribution. We obtain an independent estimate of the
exponent $\zeta$ by analyzing the scaling behavior of the moments
$\mu_r(n) \equiv \langle \vert Q_n - \langle Q_n \rangle
\vert^r \rangle$, where $r<\lambda$~\cite{momref}. For a L\'evy 
stable distribution $[\mu_r(n)]^{1/r} \sim n^{1/\zeta}$. Hence, we
plot $[\mu_r(n)]^{1/r}$ as a function of $n$ [Fig.~4b,c] and obtain an
inverse slope of $\zeta = 1.45 \pm 0.03$ --- consistent with our
previous estimate of $\zeta$~\cite{shuffle}.

Since the $q_i$ have only weak correlations (the analog of $\delta$
has the value $=0.57$), we ask how $Q_{\Delta t} \equiv
\sum_{i=1}^{N_{\Delta t}} q_i$ can show much stronger 
correlations ($\delta = 0.83$). To address this question, we note that
(i) $N_{\Delta t}$ is long-range correlated~\cite{Plerou99}, and (ii)
$P(q)$ is consistent with a L\'evy stable distribution with exponent
$\zeta$, and therefore, $N_{\Delta t}^{1/\zeta}\, P([Q_{\Delta t} -
\langle q \rangle\, N_{\Delta t}]/N_{\Delta t}^{1/\zeta})$ should,
from Eq.~(\ref{sumq}), have the same distribution as any of the $q_i$.
Thus, we hypothesize that the dependence of $Q_{\Delta t}$ on
$N_{\Delta t}$ can be separated by defining $\chi \equiv [Q_{\Delta t}
- \langle q \rangle\,N_{\Delta t}]/N_{\Delta t}^{1/\zeta}$, where
$\chi$ is a one-sided L\'evy-distributed variable with zero mean and
exponent $\zeta$~\cite{Levy,assymLevy}. To test this hypothesis, we
first analyze $P(\chi)$ and find similar asymptotic behavior to
$P(Q_{\Delta t})$ [Fig.~4d].  Next, we analyze correlations in $\chi$
and find only weak correlations [Fig.~4e,f] --- implying that the
correlations in $Q_{\Delta t}$ are largely due to those of $N_{\Delta
t}$.

An interesting implication is an explanation for the
previously-observed~\cite{Karpoff,volume} equal-time correlations
between $Q_{\Delta t}$ and volatility $V_{\Delta t}$, which is the
local standard deviation of price changes $G_{\Delta t}$. Now
$V_{\Delta t} = W_{\Delta t}\, \sqrt{N_{\Delta t}}$, since $G_{\Delta
t}$ depends on $N_{\Delta t}$ through the relation $ G_{\Delta t} =
W_{\Delta t}\, \sqrt{N_{\Delta t}}\, \epsilon\,,$ where $\epsilon$ is
a Gaussian-distributed variable with zero mean and unit variance and
$W_{\Delta t}^2$ is the variance of price changes due to all
$N_{\Delta t}$ transactions in $\Delta t$~\cite{Plerou99}. Consider
the equal-time correlation, $\langle Q_{\Delta t} \, V_{\Delta t}
\rangle$, where the means are subtracted from $Q_{\Delta t}$ and
$V_{\Delta t}$. Since $Q_{\Delta t}$ depends on $N_{\Delta t}$ through
$Q_{\Delta t} = \langle q \rangle N_{\Delta t} + N_{\Delta
t}^{1/\zeta}\,\chi$, and the equal-time correlations $\langle
N_{\Delta t} \, W_{\Delta t} \rangle$, $\langle N_{\Delta t}
\, \chi \rangle$, and $\langle W_{\Delta t} \, \chi
\rangle$ are small (correlation coefficient of the order of $\approx
0.1$), it follows that the equal-time correlation $\langle Q_{\Delta
t}\,V_{\Delta t}\rangle \propto \langle N_{\Delta t}^{3/2} \rangle -
\langle N_{\Delta t} \rangle \langle N_{\Delta t}^{1/2} \rangle$, 
which is positive due to the Cauchy-Schwartz inequality. Therefore,
$\langle Q_{\Delta t} \, V_{\Delta t} \rangle$ is large because of
$N_{\Delta t}$.

\begin{figure}
\narrowtext
\centerline{
\epsfysize=0.6\columnwidth{\rotate[r]{\epsfbox{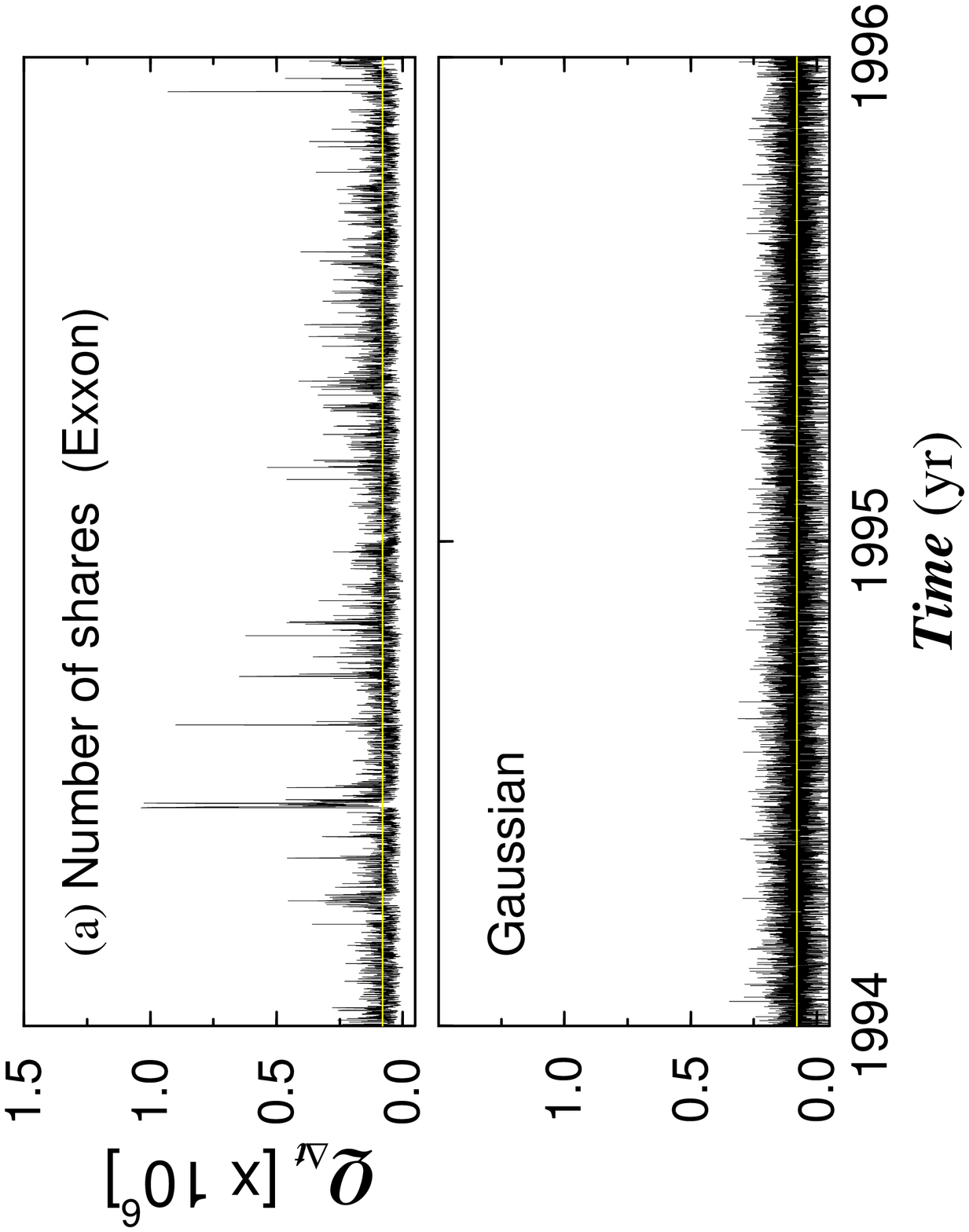}}}
}
\vspace{0.5cm}
\centerline{
\epsfysize=0.6\columnwidth{\rotate[r]{\epsfbox{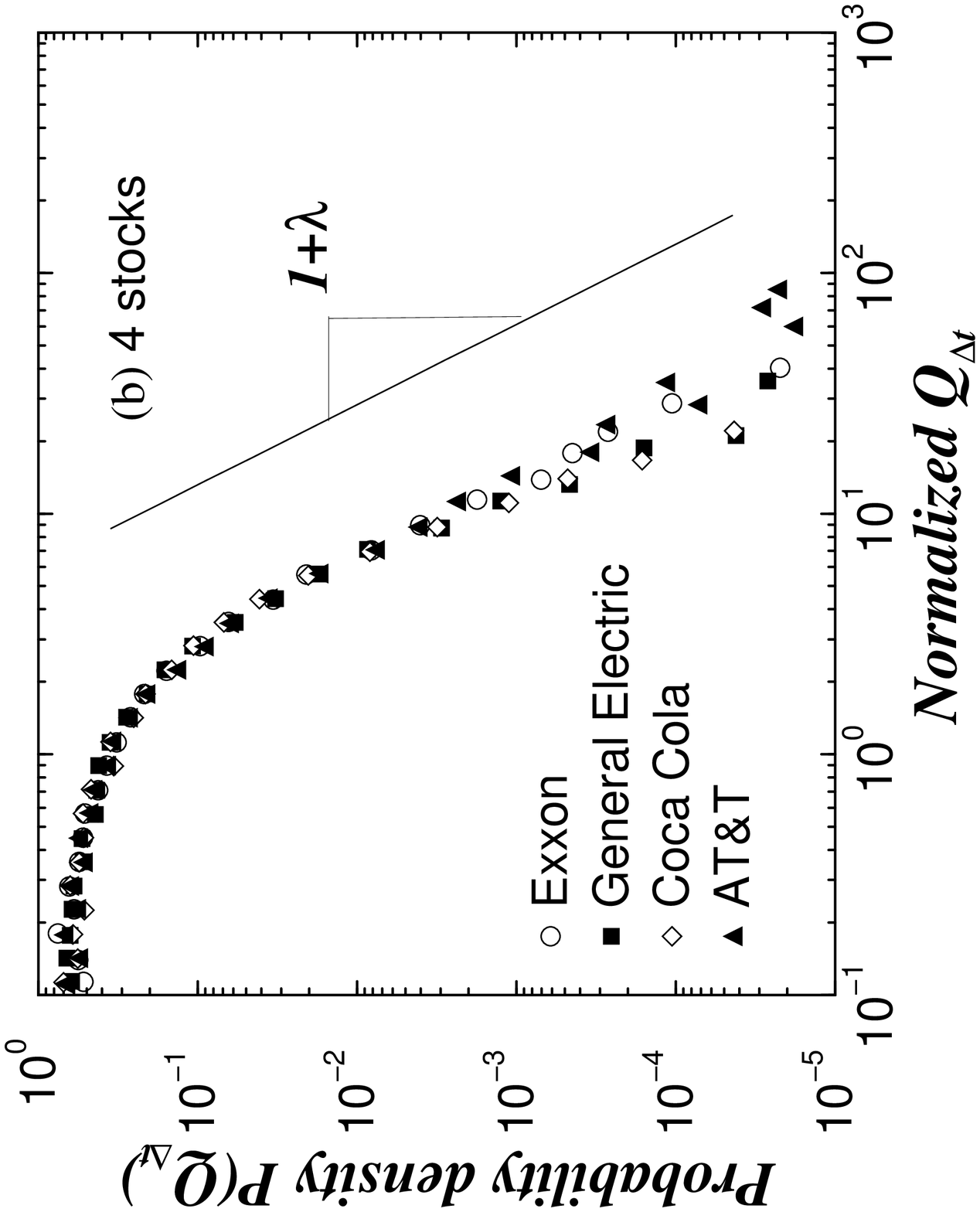}}}
}
\vspace{0.5cm}
\centerline{
\epsfysize=0.6\columnwidth{\rotate[r]{\epsfbox{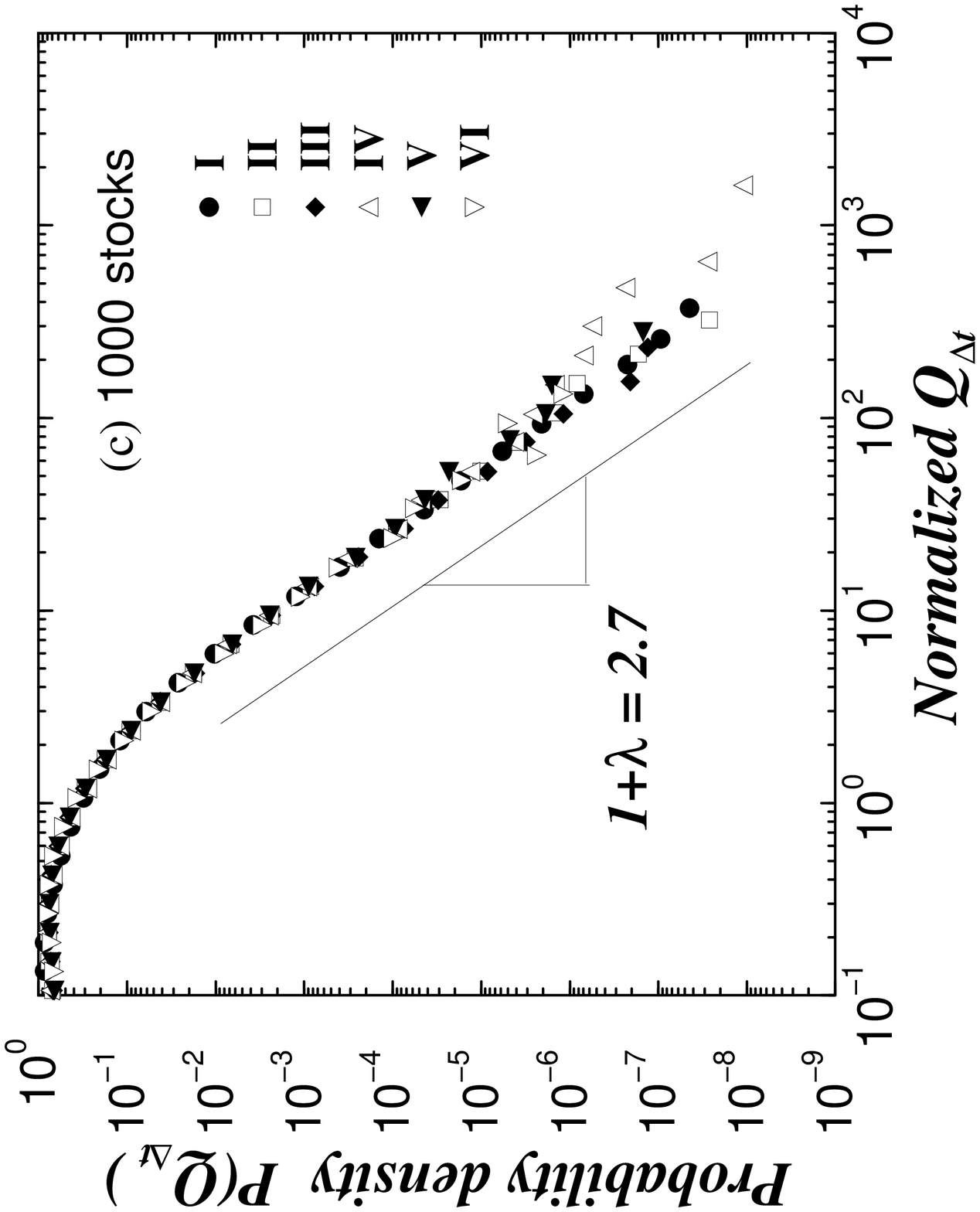}}}
}
\vspace{0.5cm}
\centerline{
\epsfysize=0.6\columnwidth{\rotate[r]{\epsfbox{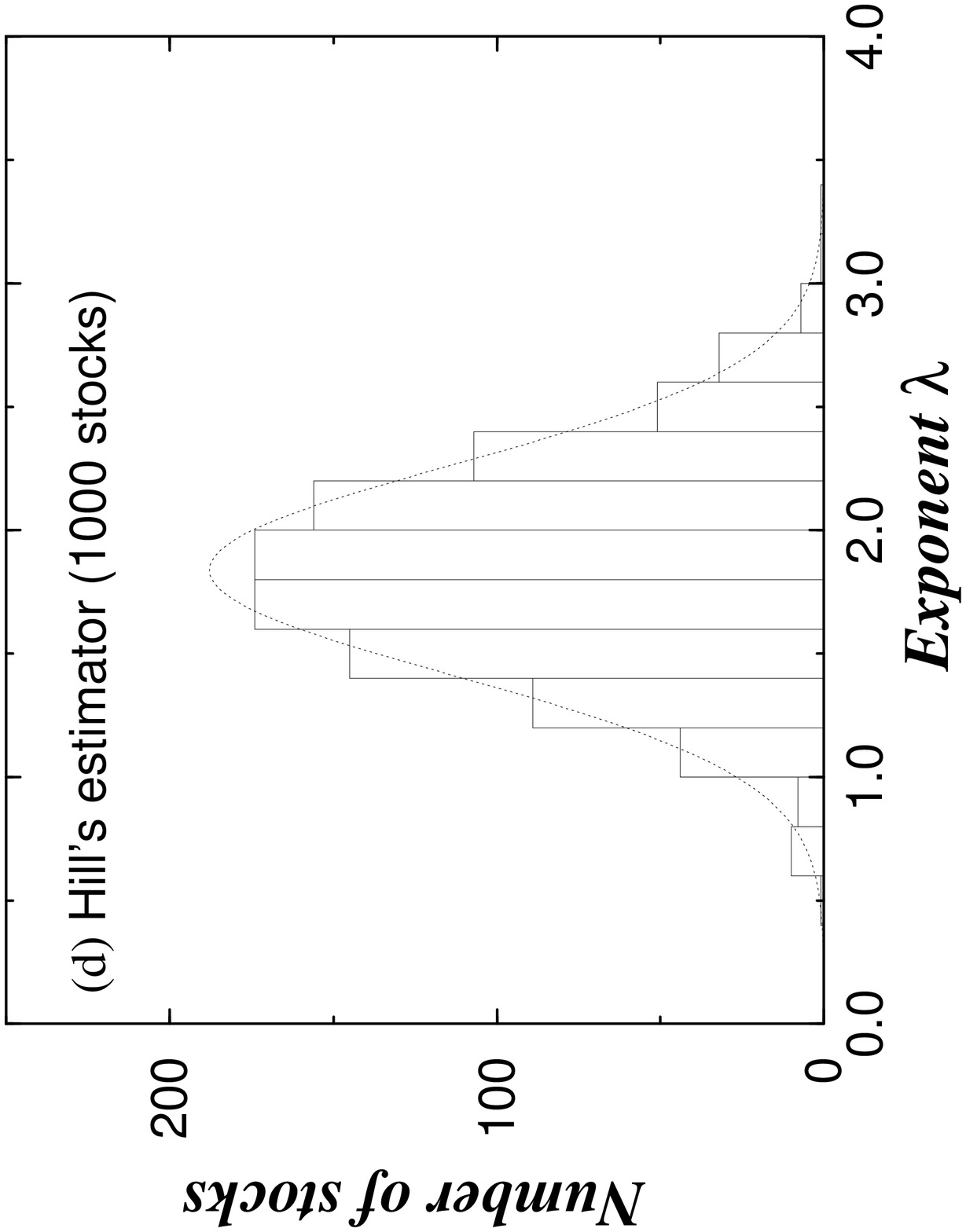}}}
}
\caption{ (a) Number of shares traded~\protect\cite{noteQ} 
for Exxon Corporation (upper panel) for an interval $\Delta t =
15$~min compared to a series of Gaussian random numbers with the same
mean and variance (lower panel). (b) Probability density function
$P(Q_{\Delta t})$ for 4 actively-traded stocks Exxon Corp., General
Electric Co., Coca Cola Corp., and A T \& T Corp., shows an asymptotic
power-law behavior characterized by an exponent $1+
\lambda$. Hill's method~\protect\cite{Hill} gives $\lambda = 1.87 \pm 0.14, 
2.10 \pm 0.17, 1.91\pm 0.20$, and $1.71 \pm 0.09$ respectively. (c)
$P(Q_{\Delta t})$ for 1000 stocks on a log-log scale.  To choose
compatible sampling time intervals $\Delta t$, we first partition the
1000 companies studied into six groups~\protect\cite{Plerou99} denoted
I - VI, based upon the average time interval between trades $\delta
t$. For each group, we choose $\Delta t > 10 \, \delta t$, to ensure
that each interval has a sufficient $N_{\Delta t}$. Thus we choose
$\Delta t = 15, 39, 65, 78, 130$ and $390$~min for groups I - VI
respectively, each containing $\approx 150$ companies. Since the
average value of $Q_{\Delta t}$ differs from one company to the other,
we normalize $Q_{\Delta t}$ by its average. Each symbol shows the
probability density function of normalized $Q_{\Delta t}$ for all
companies that belong to each group.  Power-law regressions on the
density functions of each group yield the mean value $\lambda =
1.78\pm 0.07$. (d) Histogram of exponents $\lambda_i$ for
$i=1,\dots,1000$ stocks obtained using Hill's
estimator~\protect\cite{Hill}, shows an approximately Gaussian spread
around the average value $\lambda = 1.7\pm 0.1$.}
\label{figQ}
\end{figure}

\begin{figure}
\narrowtext
\centerline{
\epsfysize=0.6\columnwidth{\rotate[r]{\epsfbox{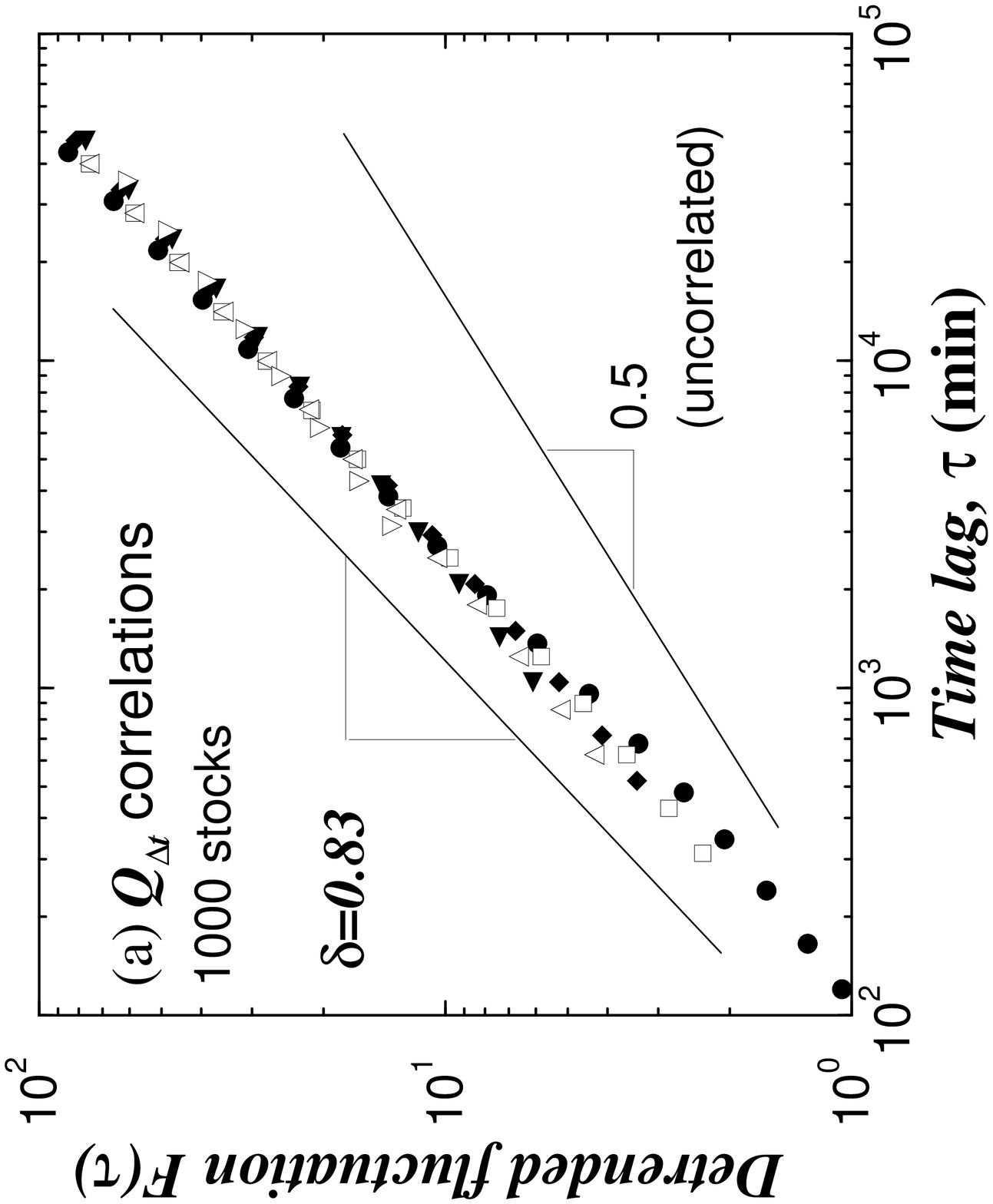}}}
}
\vspace{0.5cm}
\centerline{
\epsfysize=0.6\columnwidth{\rotate[r]{\epsfbox{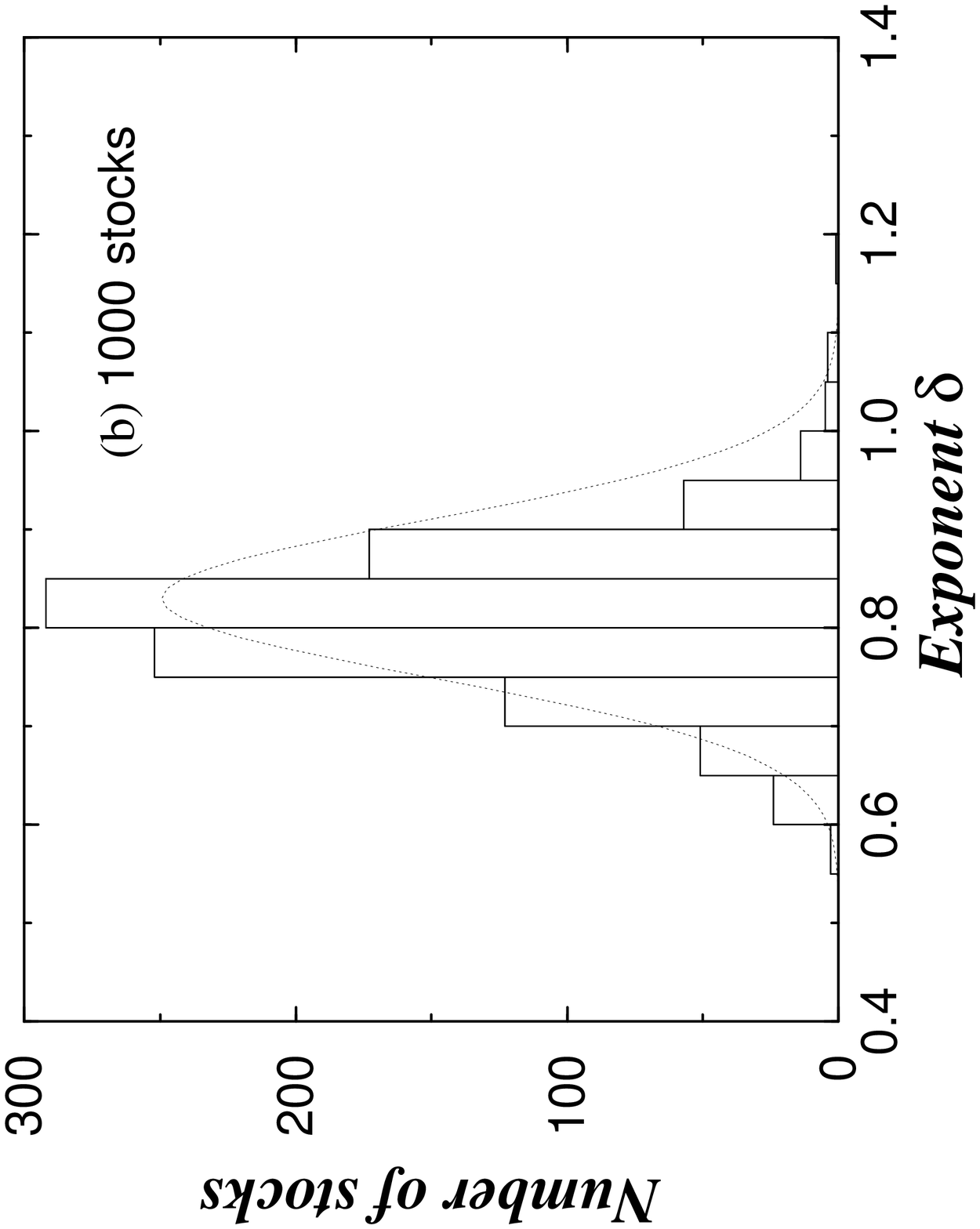}}}
}
\caption{ (a) Detrended fluctuation function $F(\tau)$ for 
$(Q_{\Delta t})^a$ for $a=0.5$~\protect\cite{diffa}, averaged for all
stocks within each group (I-VI) as a function of the time lag
$\tau$. $F(\tau)$ for a time series is defined as the $\chi^2$
deviation of a linear fit to the integrated time series in a box of
size $\tau$~\protect\cite{Peng94}. An uncorrelated time series
displays to $F(\tau) \sim \tau^{\delta}$, where $\delta = 0.5$,
whereas long-range correlated time series display values of exponent
in the range $0.5 < \delta \leq 1$. In order to detect genuine
long-range correlations, the U-shaped intraday pattern for $Q_{\Delta
t}$ is removed by dividing each $Q_{\Delta t}$ by the intraday
pattern~\protect\cite{Yanhui97}.  (b) Histogram of $\delta $ obtained
by fitting $F(\tau)$ with a power-law for each of the 1000
companies. We obtain a mean value of the exponent $0.83\pm 0.02$.}
\label{figcorrQ}
\end{figure}

\nopagebreak
\begin{figure}
\narrowtext
\centerline{
\epsfysize=0.6\columnwidth{\rotate[r]{\epsfbox{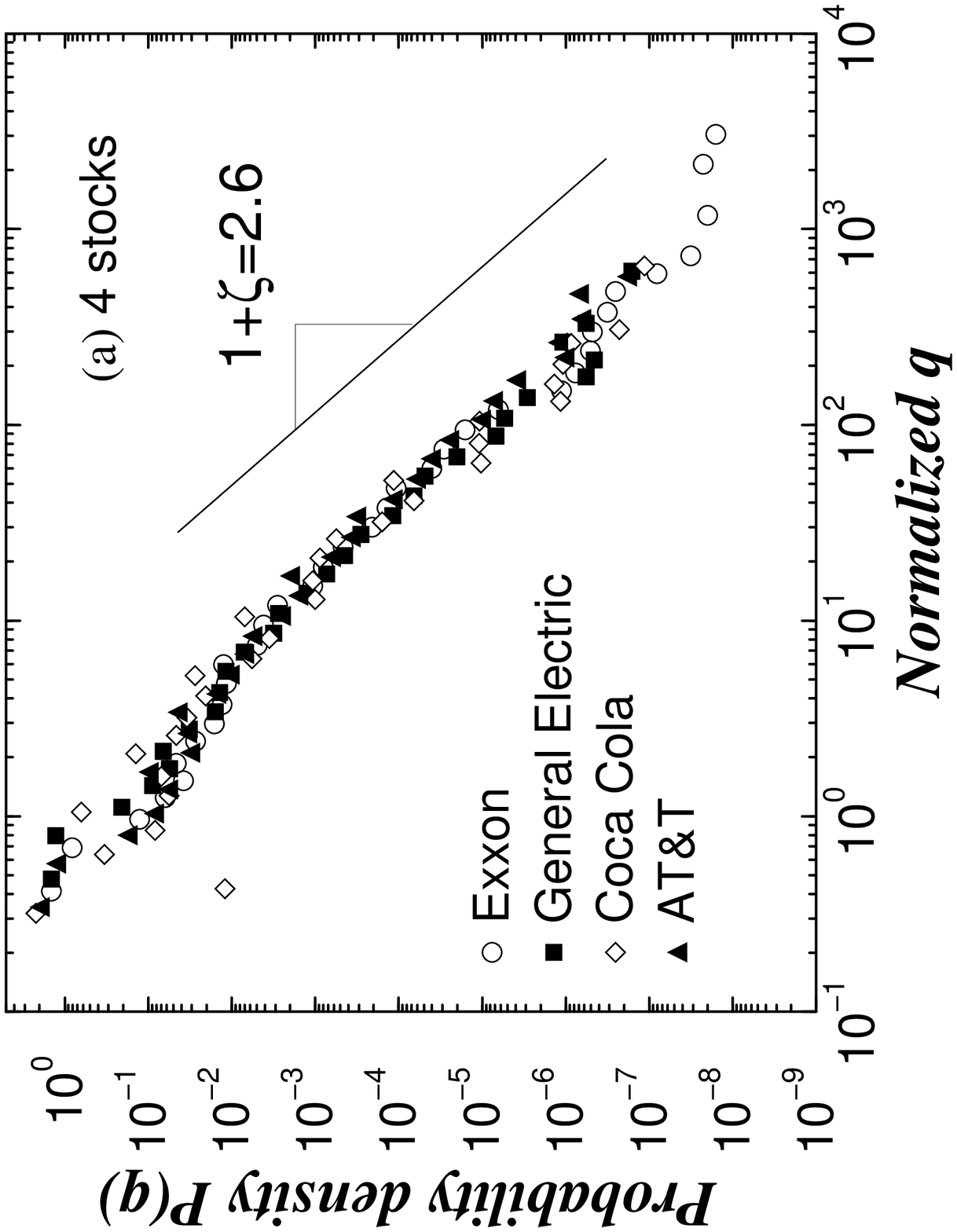}}}
}
\pagebreak
\vspace*{0.5cm}
\centerline{
\epsfysize=0.6\columnwidth{\rotate[r]{\epsfbox{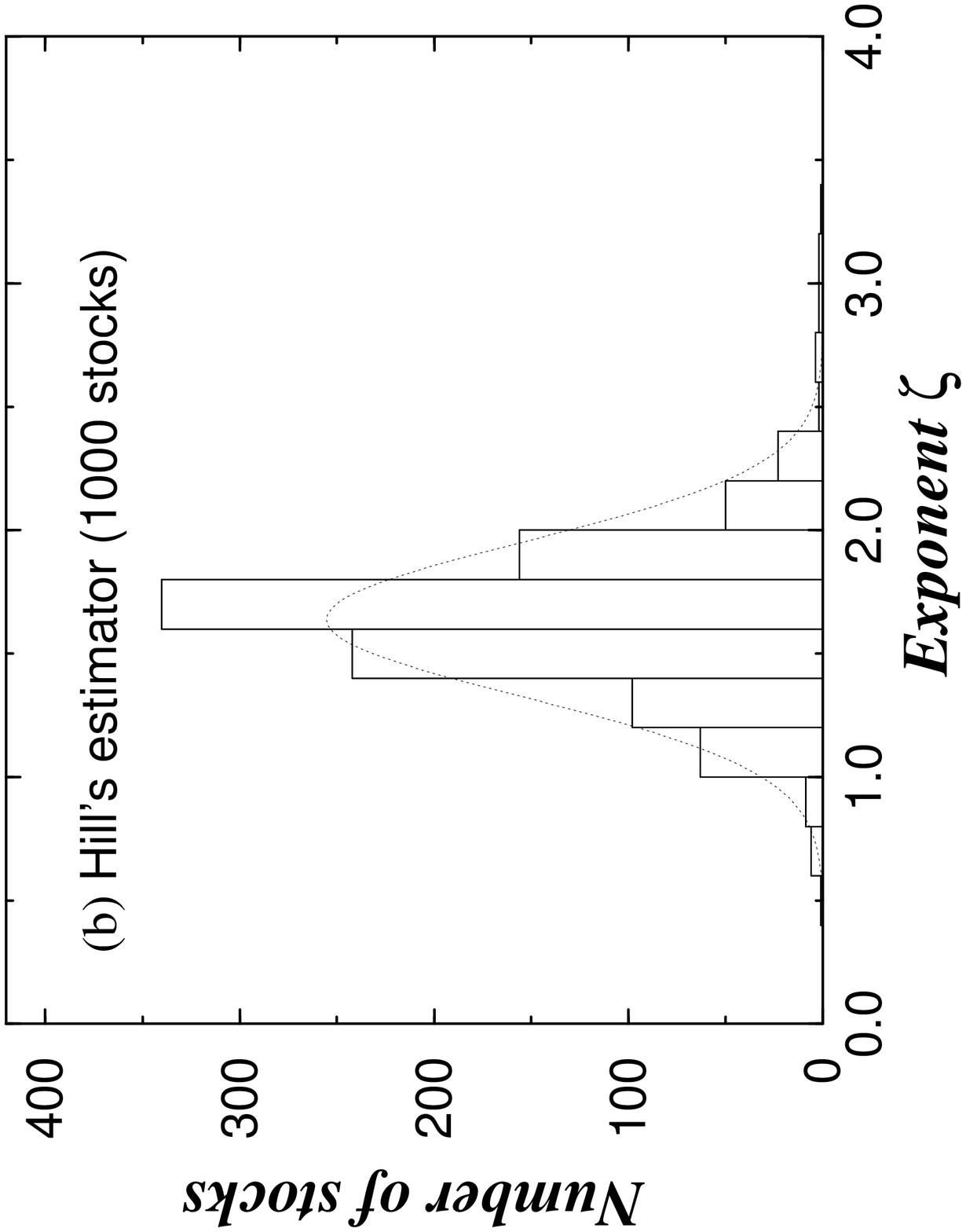}}}
}
\caption{(a) Probability density function of the number of shares $q_i$
traded, normalized by the average value, for all transactions for the
same four actively-traded stocks. We find an asymptotic power-law
behavior characterized by an exponent $\zeta$.  Fits yield values
$\zeta = 1.87\pm0.13, 1.61 \pm 0.08, 1.66\pm0.05, 1.47\pm 0.04$,
respectively for each of the 4 stocks. (b) Histogram of the values of
$\zeta$ obtained for each of the 1000 stocks using Hill's
estimator~\protect\cite{Hill}, whereby we find the average value
$\zeta = 1.53 \pm 0.07$.}
\label{figunsq}
\end{figure}

\begin{figure}
\centerline{
\epsfysize=0.6\columnwidth{\rotate[r]{\epsfbox{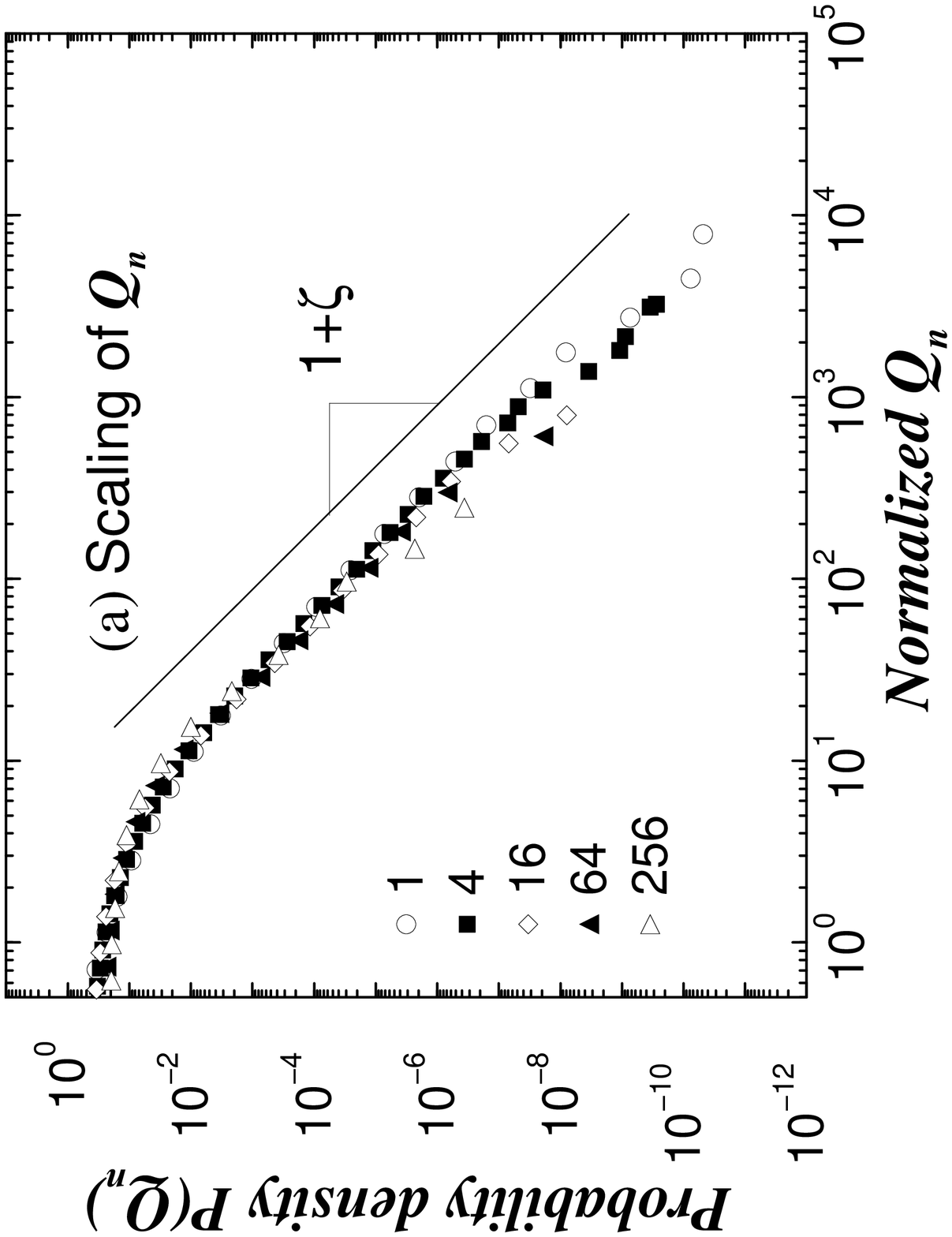}}}
}
\vspace{0.5cm}
\centerline{
\epsfysize=0.6\columnwidth{\rotate[r]{\epsfbox{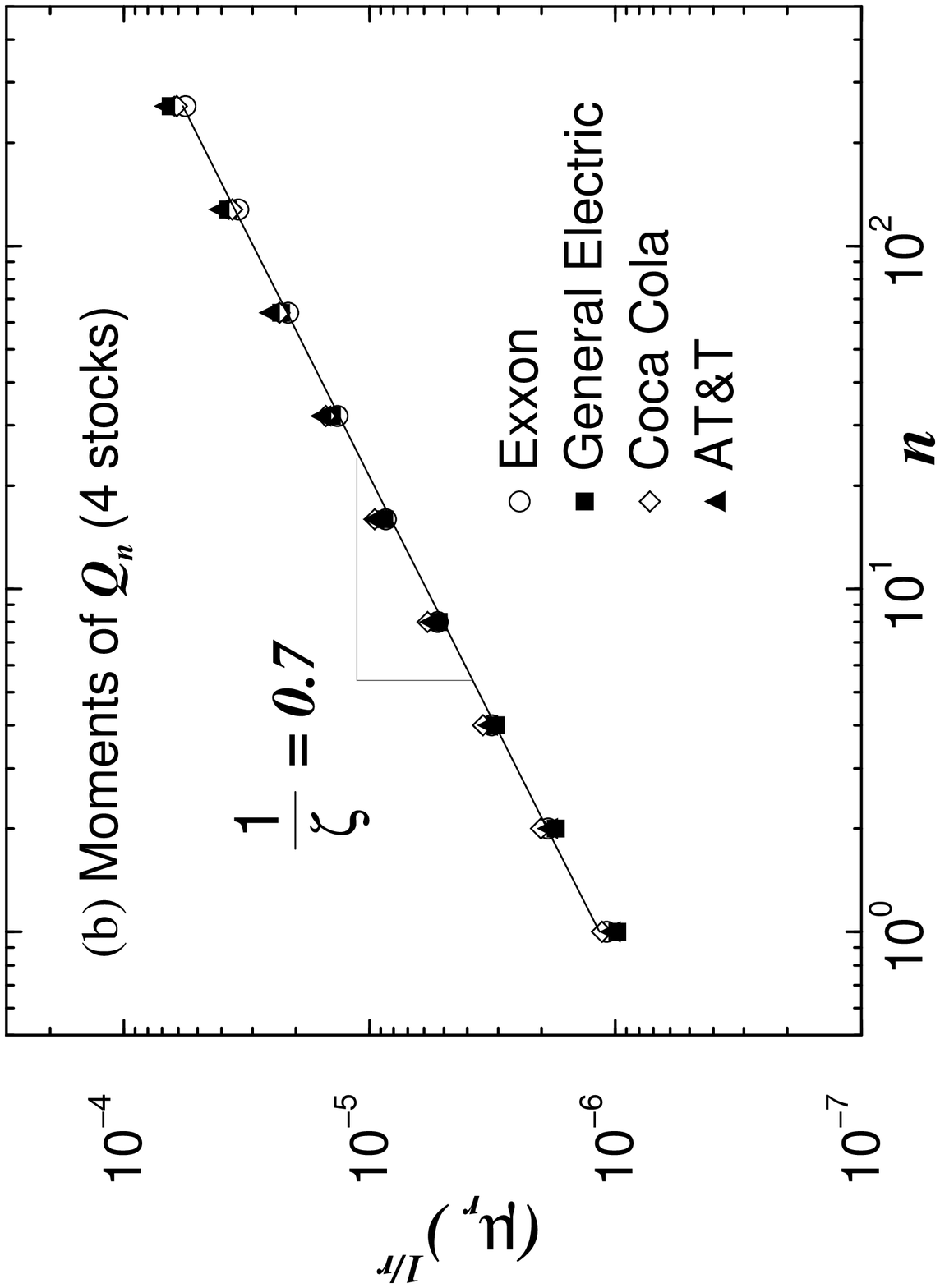}}}
}
\vspace{0.5cm}
\centerline{
\epsfysize=0.6\columnwidth{\rotate[r]{\epsfbox{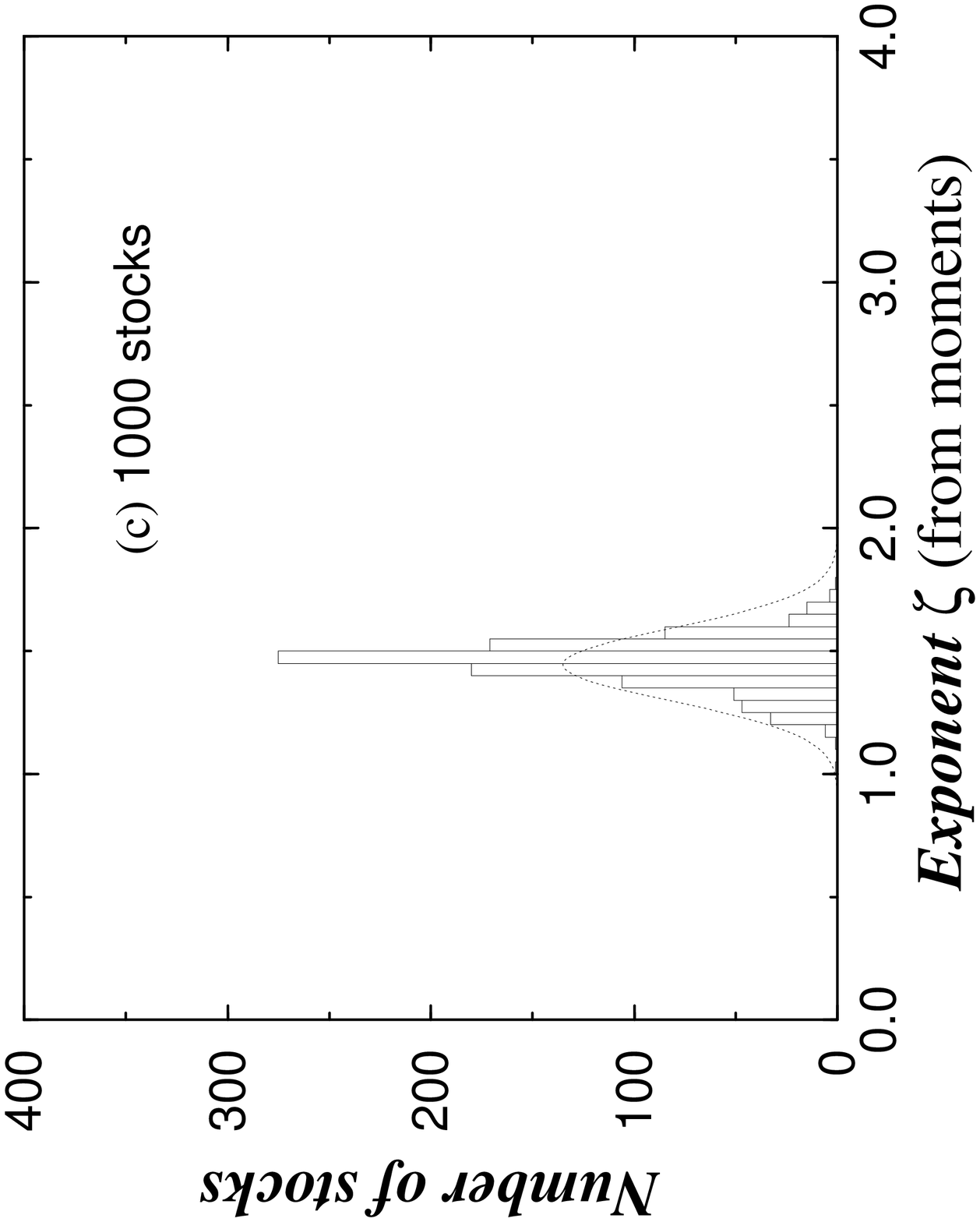}}}
}
\vspace{0.5cm}
\centerline{
\epsfysize=0.6\columnwidth{\rotate[r]{\epsfbox{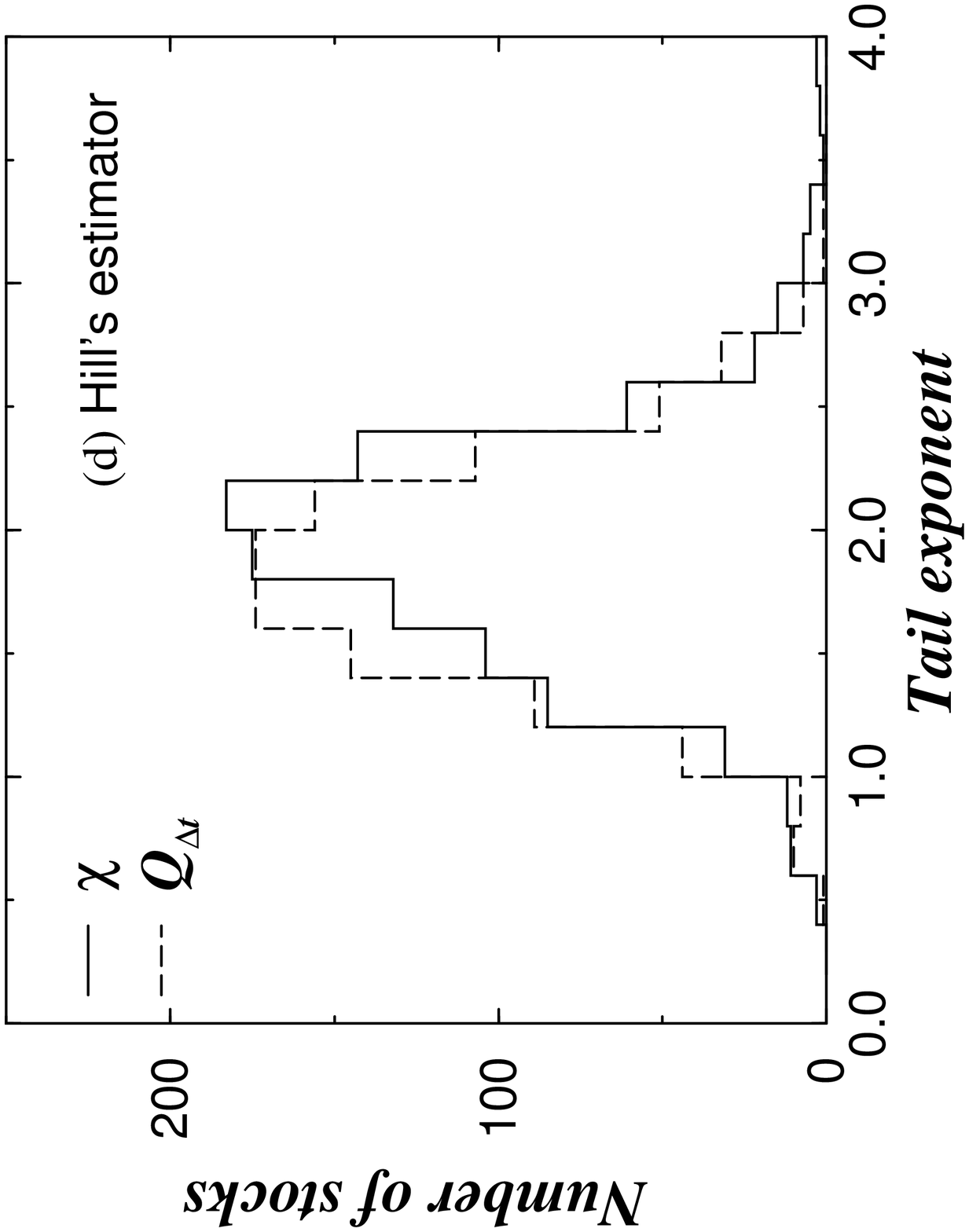}}}
}
\vspace{0.5cm}
\centerline{
\epsfysize=0.6\columnwidth{\rotate[r]{\epsfbox{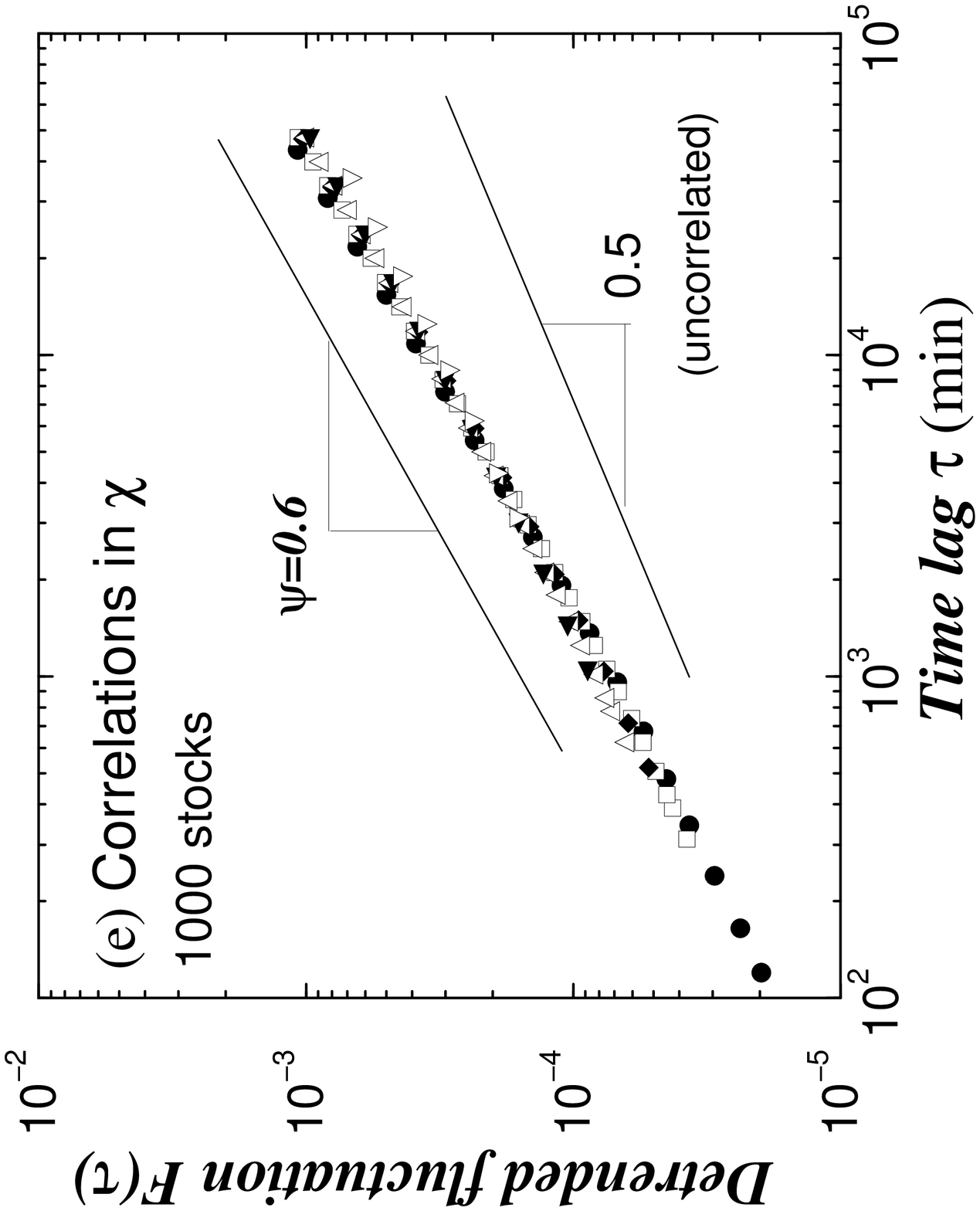}}}
}
\vspace{0.5cm}
\centerline{
\epsfysize=0.6\columnwidth{\rotate[r]{\epsfbox{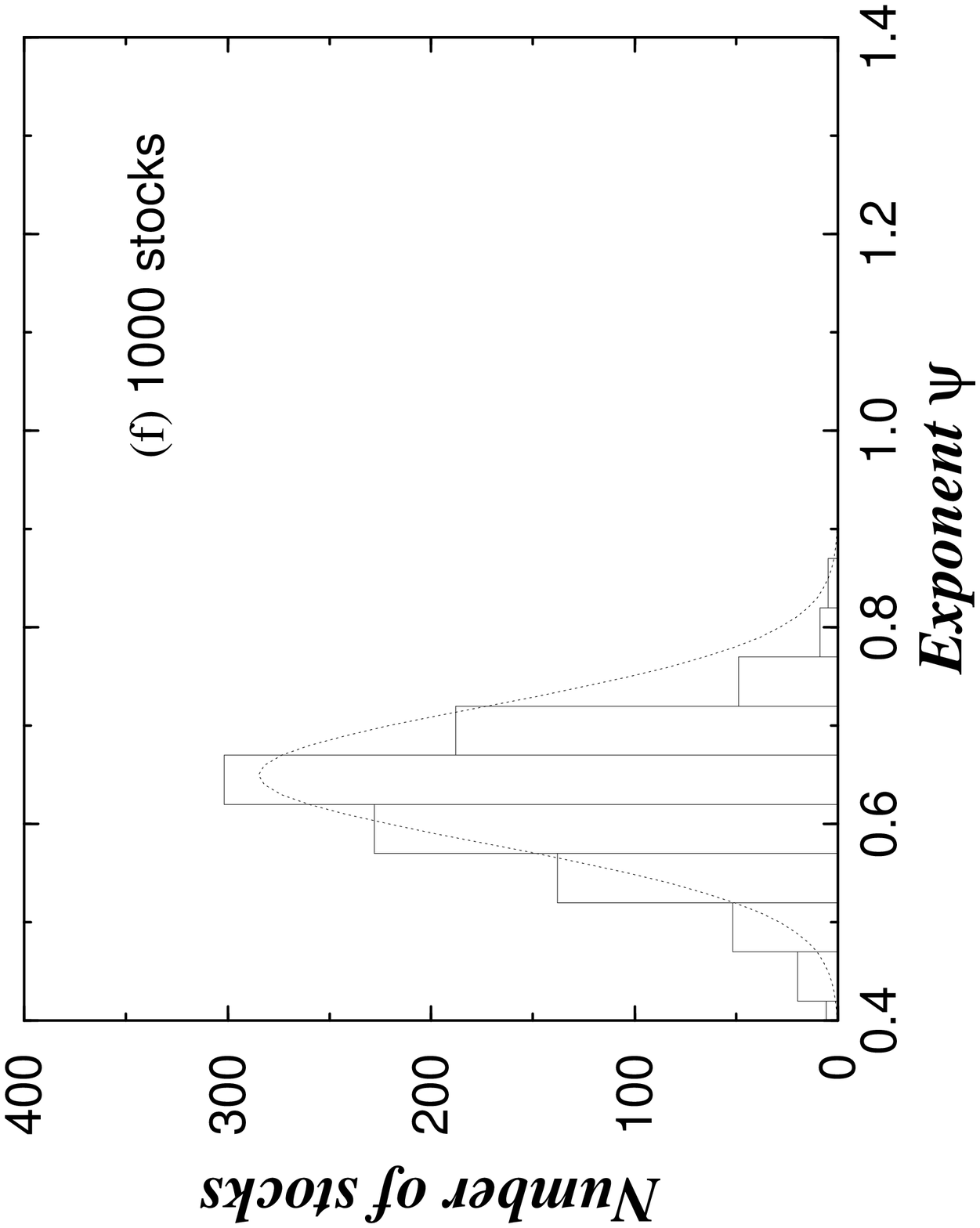}}}
}

\vspace{0.5cm}
\caption{(a) Probability distribution of $Q_n$ as a function of increasing 
$n= 1,\dots,256$ apparently retains the same asymptotic behavior. (b)
Scaling of the $r^{\rm th}$ moments $\mu_r$ with increasing $n$ for
the same four stocks. The inverse slope of this line yields an
independent estimate of the exponent $\zeta$. We obtain $\zeta = 1.43
\pm 0.02, 1.35\pm 0.03, 1.42 \pm 0.01, 1.41 \pm 0.02$
respectively. (c) Histogram of exponents $\zeta$ obtained by fitting a
power-law to the equivalent of part (b) for all 1000 stocks
studied. We thus obtain a value $\zeta = 1.45\pm 0.03$ consistent with
our previous estimate using Hill's estimator. (d) Histogram of slopes
estimated using Hill's estimator for the scaled variable $\chi\equiv
[Q_{\Delta t} - \langle q \rangle N_{\Delta t}]/ N_{\Delta
t}^{1/\zeta}$ compared to that of $Q_{\Delta t}$. We obtain a mean
value $1.7 \pm 0.1$ for the tail exponent of $\chi$, consistent with
our estimate of the tail exponent $\lambda$ for $Q_{\Delta t}$. (e)
Detrended fluctuation function $F(\tau)$ for $\chi$, where each symbol
denotes an average of $F(\tau)$ for all stocks within each group (I-VI
as in Fig.~1). (f) Histogram of detrended fluctuation exponents for
$\chi$.  We obtain an average value for the exponent $0.61 \pm 0.03$
which indicates only weak correlations compared to the value of the
exponent $\delta= 0.83\pm 0.03$ for $Q_{\Delta t}$.}
\label{figconstn}
\end{figure}

\end{multicols}


\begin{references} 
\bibitem{Mantegna95} 
J.D. Farmer, Computing in Science \& Engineering {\bf 1}, 26 (1999).

\bibitem{Lux}
T.~Lux, Applied Financial Economics {\bf 6}, 463 (1996);
P.~Gopikrishnan {\it et al.} Phys. Rev. E. {\bf 60}, 5305 (1999);
V.~Plerou {\it et al.}, {\it ibid.} {\bf 60}, 6519 (1999).

\bibitem{Yanhui97} 
Y.~Liu {\it et al.}, Phys. Rev. E {\bf 60}, 1390 (1999); M.~Lundin
{\it et al.}, in {\it Financial Markets Tick by Tick,} P.~Lequeux
(ed), p.91 (J. Wiley, New York 1999); Z. Ding {\it et al.}, J.
Empirical Finance {\bf 1}, 83 (1993).

 
\bibitem{TAQ} 
{\it The Trades and Quotes Database\/}, 24 CD-ROMs for 1994-95,
published by the New York Stock Exchange.

\bibitem{Peng94} 
C.-K.~Peng {\it et al.}, Phys. Rev. E {\bf 49}, 1685 (1994).

\bibitem{Bunde} E.~Koscielny-Bunde, {\it et al.}, Phys. Rev. Lett. 
{\bf 81}, 729 (1998); C.-K.~Peng {\it et al.}, {\it ibid.} {\bf 70}, 1343
(1993); Nature (London) {\bf 356}, 168 (1992).


\bibitem{diffa}
Here, $\kappa$ is the exponent characterizing the decay of the
autocorrelation function, compactly denoted $\langle [Q_{\Delta
t}(t)]^a [Q_{\Delta t}(t+\tau)]^a \rangle$. Values of $a$ in the range
$0.1 < a < 1$ yield $\delta$ in the range $0.75 < \delta < 0.88$ ---
consistent with long-range correlations in $Q_{\Delta t}$.


\bibitem{Levy} 
M. F. Shlesinger {\it et al.} (eds.), {\it L\'evy Flights and Related
Topics in Physics} (Springer, Berlin, 1995); C.~Tsallis, Physics World
{\bf 10}, 42 (1997); J.-P.~Bouchaud and A.~Georges, Phys. Rep. {\bf
195}, 127 (1990).


\bibitem{assymLevy} 
The general form of a characteristic function of a L\'evy stable
distribution is $ \ln \varphi(x) \equiv i\mu
x-\gamma|x|^\alpha\left[1+i\beta{x\over|x|}tg\left( {\pi\over
2}\alpha\right)\right]\,\,[\alpha\neq 1]$, where $0<\alpha<2$,
$\gamma$ is a positive number, $\mu$ is the mean, and $\beta$ is an
asymmetry parameter. The case where the parameter $\beta =1$ gives a
positive or one-sided L\'evy stable distribution.

\bibitem{momref}
The values of $\zeta$ reported are using $r=0.5$. Varying $r$ in the
range $0.2 < r < 1$ yields similar values.

\bibitem{shuffle}
To avoid the effect of weak correlations in $q$ on the estimate of
$\zeta$, the moments $[\mu_r(n)]^{1/r}$ are constructed after
randomizing each time series of $q_i$. Without randomizing, the same
procedure gives an estimate of $\zeta = 1.31 \pm 0.03$.

\bibitem{Karpoff}
J.~Karpoff, Journal of Financial and Quantitative Analysis {\bf 22},
109 (1987); C.~Jones {\it et al.}, Reviews of Financial Studies {\bf
7}, 631 (1994); A.~R.~Gallant {\it et al.}, {\it ibid.}  {\bf 5}, 199
(1992).

\bibitem{volume}
G. Tauchen and M. Pitts, Econometrica {\bf 57}, 485 (1983); T.W. Epps
and M.L. Epps, {\it ibid.} {\bf 44}, 305 (1976); P.K. Clark, {\it
ibid.} {\bf 41}, 135 (1973).

\bibitem{Plerou99} 
V.~Plerou {\it et al.}, cond-mat/9912051.

\bibitem{noteQ}
Opening trades are not shown in this plot. For all calculations, we
have normalized $Q_{\Delta t}$ by the total number of outstanding
shares in order to account for stock splits.

\bibitem{Hill}  B.~M.~Hill, Ann. Stat. {\bf3}, 1163 (1975).


\vspace{-2cm}
\end{references}
\end{document}